\documentclass[aps,prd,twocolumn,showpacs,10pt,superscriptaddress, footinbib]{revtex4-1}

\usepackage{amsmath, amsthm, amssymb,bm,relsize}
\usepackage{amsmath,autobreak}
\usepackage{graphicx}
\usepackage{color}

\usepackage{dsfont}
\newcommand\BbbGamma{\reflectbox{\rotatebox[origin=c]{180}{$\mathds L$}}}

\usepackage{caption}
\usepackage{booktabs,siunitx}
\allowdisplaybreaks
\usepackage{tabularx}
\newcolumntype{P}[1]{>{\centering\arraybackslash}p{#1}}
\usepackage{multirow}
\usepackage{slashed}
\usepackage{array}

\usepackage{endnotes}
\let\footnote=\endnote

\usepackage[caption=false]{subfig}
\definecolor{urlblue}{rgb}{0.2,0.4,0.7}
\definecolor{citegreen}{rgb}{0,0.6,0.2}
\definecolor{linkred}{rgb}{0.9,0.2,0.1}
\usepackage{hyperref}
\hypersetup{colorlinks=true, citecolor=citegreen, linkcolor=blue,urlcolor=urlblue}

\def\la{\leftarrow}

\allowdisplaybreaks

\begin{document}

\title{NNLO QCD\texorpdfstring{$\otimes$}{}QED corrections to unpolarized and polarized SIDIS}

\author{Saurav Goyal}
\email{sauravg@imsc.res.in}
\affiliation{The Institute of Mathematical Sciences,  Taramani, 600113 Chennai, India}
\affiliation{Homi Bhabha National Institute, Training School Complex, Anushakti Nagar, Mumbai 400094, India}
\author{Roman N. Lee}
\email{r.n.lee@inp.nsk.su}
\affiliation{Budker Institute of Nuclear Physics, 630090, Novosibirsk, Russia}
\author{Sven-Olaf Moch}
\email{sven-olaf.moch@desy.de}
\affiliation{II. Institute for Theoretical Physics, Hamburg University, D-22761 Hamburg, Germany} 
\author{Vaibhav Pathak}
\email{vaibhavp@imsc.res.in}
\affiliation{The Institute of Mathematical Sciences, Taramani, 600113 Chennai, India}
\affiliation{Homi Bhabha National Institute, Training School Complex, Anushakti Nagar, Mumbai 400094, India}
\author{V. Ravindran}
\email{ravindra@imsc.res.in}
\affiliation{The Institute of Mathematical Sciences, Taramani, 600113 Chennai, India}
\affiliation{Homi Bhabha National Institute, Training School Complex, Anushakti Nagar, Mumbai 400094, India}

\date{\today}

\begin{abstract}
We present the first computation of next-to-next-to-leading order (NNLO) pure QED and mixed QCD$\otimes$QED corrections to unpolarized and polarized  semi-inclusive deep-inelastic scattering (SIDIS). 
Building on our previous NNLO QCD results, these corrections are crucial for improving the theoretical precision. 
The coefficient functions are derived within the QCD factorization framework using dimensional regularization, with consistent renormalization and mass factorization. 
A detailed phenomenological analysis shows that the NNLO QED and QCD$\otimes$QED terms enhance perturbative stability and reduce scale uncertainties. 
These results are essential for high-precision SIDIS predictions at future facilities such as the Electron-Ion Collider.
\end{abstract}

\maketitle
Semi-inclusive deep-inelastic scattering (SIDIS), characterized by the detection of a final-state hadron along with the scattered lepton, serves as a powerful tool for probing the internal structure of both incoming as well as outgoing hadrons.  Unlike inclusive DIS, SIDIS provides access to both parton distribution functions (PDFs) and fragmentation functions (FFs), offering a more detailed picture of the nucleon's partonic structure and an opportunity for extracting FFs.  
QCD factorization separates the perturbatively calculable coefficient functions (CFs) from the nonperturbative PDFs and FFs and allows for the SIDIS cross section to be expressed as a convolution of these ingredients, PDFs, FFs, and CFs. 
The CFs are computed order-by-order in perturbation theory and are known up to next-to-next-to-next-to-leading order (N$^3$LO) for inclusive DIS~\cite{Moch:2004xu,Vermaseren:2005qc,Moch:2008fj,Blumlein:2022gpp}, 
while for SIDIS, the CFs at next-to-next-to-leading order (NNLO) in QCD have been completed recently~\cite{Goyal:2023zdi,Bonino:2024qbh,Bonino:2024wgg,Goyal:2024tmo,Bonino:2025tnf,Bonino:2025qta,Bonino:2025bqa}.
Building upon these developments, we present the first results on the impact of NNLO QED and mixed QCD$\otimes$QED contributions on unpolarized and polarized SIDIS. These contributions are essential for enhancing the precision of theoretical predictions, reducing  uncertainties resulting from renormalisation and factorization scales, and are particularly relevant for upcoming high-precision experiments such as the Electron-Ion Collider (EIC)~\cite{AbdulKhalek:2021gbh} at BNL.

The reaction $l(k_l) + H(P) \rightarrow  l({k}'_l) + H'(P_H) + X$ defines the SIDIS process, where $k_l$, ${k}'_l$ ($P$, $P_H$) are the momenta of the incoming and outgoing leptons (hadrons), and the virtuality of the exchanged photon is given by $Q^2=-q^2$, with $q= k_l - {k}'_l$ . 
For unpolarized (polarized) SIDIS, the hadronic cross section is described by the structure functions (SFs) $F_{1,2,3}$ ($g_{1,\cdots,5}$).  
The dominant quantum corrections to the CFs of $F_{1}$, $F_{2}$ and $g_{1}$ result from QCD and they have been extensively studied up to NNLO, see ~\cite{Altarelli:1979kv,Furmanski:1981cw,Cacciari:2001cw,Anderle:2012rq,Anderle:2013lka,Abele:2021nyo,Abele:2022wuy}.
The SFs of SIDIS also receive contributions from electroweak interactions via vector bosons such as $Z, W^\pm$, as well as from additional photons, with the simplest corrections arising from photons—namely, quantum electrodynamic (QED) effects. 
There have been several studies where QED and mixed QCD$\otimes$QED corrections are considered for observables in DIS, see~\cite{deFlorian:2015ujt,deFlorian:2016gvk,deFlorian:2025yar,AH:2019xds,AH:2019pyp,Cieri:2018sfk,deFlorian:2018wcj,Cieri:2020ikq,Kilgore:2011pa}.  
In this article, we incorporate QED as well as mixed QCD$\otimes$QED corrections to $F_{1}$, $F_{2}$ and $g_{1}$ at NNLO accuracy, thus providing a more precise theoretical description of SIDIS observables.

The perturbative expansion of the SIDIS cross section in both the strong and electromagnetic couplings is given by
\begin{align}
{\mathbb \sigma} &= {\mathbb \sigma}^{(0,0)} + a_s\, {\mathbb \sigma}^{(1,0)} + a_e\, {\mathbb \sigma}^{(0,1)} 
\nonumber\\
&
+ a_s^2\, {\mathbb \sigma}^{(2,0)} 
 + a_e^2\, {\mathbb \sigma}^{(0,2)} + a_s a_e\,{\mathbb \sigma}^{(1,1)} 
+ \cdots
\label{eq:crosssection}
\end{align}
with $a_s=\alpha_s/(4 \pi)=g_s^2/(16 \pi^2)$ and $a_e=\alpha_e^2/{4 \pi}=e^2/(16 \pi^2)$ 
in terms of strong coupling $g_s$ and electric charge $e$.  The cross sections ${\mathbb \sigma}^{(m,n)}$ denote the contributions at order $\mathcal{O}(\alpha_s^m,\alpha_e^n)$. In this work, we focus on the pure QED corrections at $\mathcal{O}(\alpha_e^2)$ as well as the mixed QCD$\otimes$QED contributions at $\mathcal{O}(\alpha_s \alpha_e)$ to the SFs $F_1$, $F_2$ and $g_1$.

The differential SIDIS cross section, 
\begin{align}
    \frac{d^3\sigma}{dxdydz}\!&=\!\frac{4\pi  \alpha_e^{2}}{Q^2} \left[y F_1(x,z,Q^2) + \frac{(1-y)}{y} F_2(x,z,Q^2)\right]
    \, ,
\end{align}
depends on the space-like momentum transfer $Q^2$, the Bjorken variable $x=Q^2/(2 P\cdot q)$, the inelasticity $y={P\cdot q}/{P\cdot k_l}$, and the scaling variable $z={P\cdot P_H}/{P\cdot q}$ for the fraction of the initial energy transferred to the final-state hadron.  Since we restrict ourselves to photon mediated SIDIS, the SF $F_3$ does not contribute.
Similarly, the spin-dependent cross section is found to be
\begin{align}
\frac{d^3\Delta \sigma}{dxdydz} = \frac{4\pi\alpha_e^2}{Q^2} \big(2-y\big)g_1 (x,z,Q^2)\, .
\end{align}
Note that $g_2$ does not contribute since we restrict ourselves to longitudinally polarized hadron in the initial state.

Collinear factorization in the QCD improved parton model allows to express the SFs as convolution of PDFs, $(\Delta) f_{a/\text{P}}(x_1,\mu_F^2)$ and FFs, $D_{\text{H}'/b}(z_1,\mu_F^2)$ 
with CFs, $(\Delta ){\cal C}_{I,ab}(x/x_1,z/z_1,Q^2,\mu_F^2)$:
\begin{align}
\label{eq:SFdef}
F_I &= x^{I-1}\sum_{a,b= q,\overline{q},g,\gamma}\int_x^1 \frac{dx_1}{x_1} f_{a/\text{P}}(x_1,\mu_F^2) 
\nonumber \\
    &
 \times\int_z^1 \frac{dz_1}{z_1} D_{\text{H}'/b}(z_1,\mu_F^2) 
 { \cal C}_{I,ab}\left(\frac{x}{x_1},\frac{z}{z_1},Q^2,\mu_F^2\right)
\, .
\end{align}
The corresponding expression for $g_1$ will involve polarized PDFs $\Delta f_{a/\text{P}}$ and spin-dependent CFs denoted by $\Delta {\cal C}_{1,ab}$ (we set $I=1$).
Here $x_1$ is the incoming parton's momentum fraction  with respect to the hadron $H(P)$, i.e.\ $p_a = x_1 P$ 
and $z_1$ is the momentum fraction of the final state parton $b$  carried away by the outgoing hadron $H'(P_H)$, i.e.\ $P_H = z_1 p_b$. 
The above expression is subject to summation over the initial state partons along with the photon, collectively denoted by `$a$', from the incoming hadron and the final state partons that include photons, denoted by `$b$' that fragment into the observed hadron. 
The CFs can be computed in a perturbative expansion in powers of $a_s$ and $a_e$, 
cf.~eq.~(\ref{eq:crosssection}),
\begin{eqnarray}
\label{eq:as-ae-exp}
{(\Delta)\cal C}_{I,ab} &=& \sum_{i,j=0}^\infty\, a_s^i(\mu_R^2)\, a_e^j(\mu_R^2)\, {(\Delta)\cal C}_{I,ab}^{(i,j)}(\mu_R^2)
\, ,
\end{eqnarray}
with $\mu_F$, $\mu_R$ the factorization and renormalisation scales.
The computation of the CFs starts from the parton level cross sections denoted by $d(\Delta)\hat{\sigma}_{I,ab}$,
\begin{eqnarray}
\label{eq:partonij-crs}
d(\Delta)\hat{\sigma}_{I,ab} \!\!\! &= \! \frac{1}{4\pi} \int \text{dPS}_{X+b}~ \overline{\Sigma} \overline{\left|(\Delta){M}_{ab}\right|}^2 \,
\! \! \delta\left(\frac{z}{z_1} - \frac{p_a \cdot p_b}{p_a \cdot q}\right) ,\quad
\end{eqnarray}
where the squared amplitude is defined as $\overline{|(\Delta) {M}_{ab}|}^2 = (\Delta)\mathcal{P}_I^{\mu\nu} |(\Delta)M_{ab}|^2_{\mu\nu}$ with 
suitable projectors $(\Delta)\mathcal{P}_I^{\mu\nu} $ for the SF under consideration.  
The matrix elements $M_{ab}$ denote scattering amplitudes for the parton level subprocess  $a(p_a) + \gamma^*(q) \rightarrow b(p_b) + X$  with QCD, QED, and mixed QCD$\otimes$QED corrections taken into account.  
 For spin-independent squared amplitudes,  $\overline{\Sigma}$ 
represents the average over initial and the sum over final spins/polarizations, and color quantum numbers for partons. 
For spin-dependent cross sections,
instead of averaging over the initial polarizations, we take their difference to account for the parton spin dependence.
The integration measure $\text{dPS}_{X+b}$ is the phase space for the fragmenting parton $b$ and the remaining final-state particles $X$.
 
The spin-dependent squared amplitudes, $\overline{|\Delta  {M}_{ab}|}^2$, in eq.~(\ref{eq:partonij-crs}), involve either the Dirac matrix $\gamma_5$ or the Levi-Civita tensor $\epsilon_{\mu\nu\rho\sigma}$, arising from helicity-dependent quark (or antiquark) wavefunctions or gluon(photon) polarization states, respectively; see, e.g.,~\cite{Zijlstra:1993sh}. Since both $\gamma_5$ and $\epsilon_{\mu\nu\rho\sigma}$ are intrinsically four-dimensional objects, a consistent prescription is required to define them in dimensional regularization with $d = 4 + \varepsilon$ dimensions.

Several schemes exist for this purpose, though none fully preserves the chiral Ward identities. In this work, we adopt Larin’s scheme~\cite{Larin:1993tq} to define $\gamma_5$ in $d$-dimensions as
\begin{align}
\slashed{p}_a \gamma_5 = -\frac{i}{6}~\epsilon_{\mu\nu\sigma\lambda}~p_a^\mu \gamma^\nu \gamma^\sigma \gamma^\lambda
\, ,
\end{align}
which provides a consistent framework for handling $\gamma_5$ within dimensional regularization,  where the product of two Levi–Civita tensors is expressed as a determinant of Kronecker deltas in $d$-dimensions.
The QCD contributions computed in Larin’s scheme can be translated into the \(\overline{\text{MS}}\) scheme by applying a finite renormalization. The corresponding renormalization constant~\cite{Matiounine:1998re,Ravindran:2002na,Ravindran:2003gi,Moch:2014sna} restores the axial Ward identity and simultaneously transforms the spin-dependent PDFs into the \(\overline{\text{MS}}\) scheme, see also~\cite{Goyal:2024emo,Bonino:2025bqa} for a detailed discussion in SIDIS.
Currently, no analogous renormalization scheme exists for converting the QED and mixed QCD$\otimes$QED contributions from Larin’s scheme to the \(\overline{\text{MS}}\) scheme. However, using Abelianisation we determine the finite renormalisation constants for QED and mixed QCD$\otimes$QED from those of QCD.  The results for $\Delta \mathcal {C}_{1,ab}$ upto NNLO in the \(\overline{\text{MS}}\)-scheme, thus obtained are included in the ancillary file.

Beyond leading order (LO) in perturbation theory, both ultraviolet (UV) and infrared (IR) divergences from soft and collinear partons appear in the computation of $d(\Delta) \hat \sigma_{I,ab}$. 
The UV divergences are removed by renormalisation of the bare couplings $\hat a_s$ and $\hat a_e$ at the scale $\mu_R$.  
Let $Z_{a_c}$ denote the renormalisation factor for the coupling $a_c$, where $c = s, e$ corresponds to QCD and QED. 
The bare couplings $\hat{a}_c$ are related to the renormalised ones as 
\begin{equation}
\hat{a}_c S_\varepsilon = a_c(\mu_R^2)\, Z_{a_c}\left(a_s(\mu_R^2), a_e(\mu_R^2\right), \varepsilon)\, \left( \frac{\mu^2}{\mu_R^2} \right)^{\varepsilon/2} .
\label{eq:bare_coupling}
\end{equation}
Here $\mu$ is an arbitrary
mass scale introduced to make $\hat a_c$ dimensionless
in $d$-dimension
and $S_\varepsilon = \exp\big[\frac{\varepsilon} {2}(\gamma_E - \ln 4\pi)\big]$ with $\gamma_E$ the Euler–Mascheroni constant. 
The $Z_{a_c}$ factors satisfy renormalisation group equations given by  
\begin{align}
\mu_R^2 \frac{d}{d \mu_R^2} \ln Z_{a_c} &=-\frac{1}{a_c(\mu_R^2)} \beta_{a_c}\left(a_s(\mu_R^2), a_e(\mu_R^2)\right)\, ,
\label{rg_bt}
\end{align}
where the $\beta$ functions admit expansions
$\beta_{a_s} = - \sum_{i,j=0}^\infty \beta^{(s)}_{ij} \, a_s^{i+2} a_e^j$
and
$\beta_{a_e} = - \sum_{i,j=0}^\infty \beta^{(e)}_{ij} \, a_e^{j+2} a_s^i$, with the coefficients $\beta^{(s)}_{ij}$ and $\beta^{(e)}_{ij}$ 
accounting for contributions from pure QCD, pure QED, and mixed QCD$\otimes$QED interactions. 
To NNLO accuracy the relevant $\beta$-function coefficients read 
$\beta^{(s)}_{00} = \frac{11}{3} C_a - \frac{4}{3} T_f n_f$
and 
$\beta^{(e)}_{00} = -\frac{4}{3} \Big(N\sum_q^{n_f}  e_q^2 +\sum_l^{n_L}  e_l^2\Big)$. 
Here, $C_a = N$ is the quadratic Casimir in the adjoint representation of the color SU($N$) and the one for the fundamental representation is denoted by 
$C_f = \frac{N^2 - 1}{2N}$.
The trace normalization factor for the fundamental representation is
$T_f = \frac{1}{2}$ and $n_f$, $n_L$ are the number of active quark and lepton flavors respectively. $e_q$ and $e_l$ 
denote the electric charges of the quark $q$ and the lepton $l$ respectively.

The soft IR divergences cancel among virtual and real emission processes when all the
degenerate states are summed thanks to the KLN theorem \cite{Kinoshita:1962ur,Lee:1964is}, except for the collinear ones associated with the initial-state and the final-state fragmenting partons.
These residual divergences are factorised into (un)polarized PDFs and FFs following the standard QCD mass factorization procedure. 
In the QED and QCD$\otimes$QED framework, this procedure extends to include the photon and charged leptons in the evolution equations. The partonic cross sections in eq.~(\ref{eq:partonij-crs}) are factorised into space-like Altarelli-Parisi (AP) splitting kernels $ (\Delta)\Gamma_{c\leftarrow a} $ for 
(polarized) unpolarized PDFs and time-like unpolarized AP splitting kernels $\tilde \Gamma_{b\leftarrow d} $ for FFs, which contain all $ 1/\varepsilon $ collinear singularities, and the finite CFs $(\Delta)\mathcal{C}_{I,cd} $. 
This yields the mass factorization formula at scale $ \mu_F $,
\begin{eqnarray}
\label{eq:massfactse}
{\lefteqn{
(\Delta) \hat  {\cal C}_{I,ab}(x,z,\varepsilon) \,=\,}}
\nonumber \\
&& 
(\Delta) \Gamma_{c\leftarrow a}(x,\varepsilon) \otimes
(\Delta)  {\cal C}_{I,cd}(x,z,\varepsilon)
\, \tilde \otimes \,
\tilde\Gamma_{b\leftarrow d}(z,\varepsilon)\, ,
\qquad
\end{eqnarray}
where  the convolutions  $\otimes$,  $\tilde \otimes$ run over the respective momentum fractions. 
For the QCD$\otimes$QED interactions, the AP kernels have been extended to include pure QED and mixed QCD$\otimes$QED corrections up to two-loop order as detailed in~\cite{deFlorian:2015ujt,deFlorian:2016gvk,deFlorian:2023zkc,deFlorian:2025yar}.

Beyond LO, the contributions to $d (\Delta)\hat{\sigma}_{1,ab}$  in eq.~(\ref{eq:partonij-crs}) can be classified into virtual (V) and real (R) corrections
at next-to-leading order (NLO) and 
virtual-virtual (VV), real-virtual (RV), and real-real (RR) corrections at NNLO. 
The V (VV) contributions arise from one-(two-)loop virtual corrections to the Born process including combined QCD and QED corrections. 
The R, RV and RR contributions incorporate all relevant real emission processes and their interference terms involving QCD, QED, and mixed QCD$\otimes$QED dynamics.  
Feynman diagrams are generated using \texttt{QGRAF}~\cite{Nogueira:1991ex} and algebraic manipulations, including the application of Feynman rules, Dirac algebra, Lorentz contractions, and color and charge factor simplifications, are performed with in-house routines written in \texttt{FORM}~\cite{Kuipers:2012rf,Ruijl:2017dtg}.
\begin{table}[ht]
\centering
\smaller
\renewcommand{\arraystretch}{1.3}
\begin{tabular}{|c|l|l|}
\hline
\textbf{Order} & \textbf{Process} & \textbf{Contribution type} \\
\hline
\multirow{3}{*}{NLO}
& $q + \gamma^* \to q$ (1-loop) & QCD, QED \\
& $q + \gamma^* \to q + g(\gamma)$ & QCD (QED) \\
& $g(\gamma) + \gamma^* \to q + \bar{q}$ & QCD (QED) \\
\hline
\multirow{12}{*}{NNLO}
& $q + \gamma^* \to q$ (2-loop) & QCD, QED, QCD$\otimes$QED \\
& $q + \gamma^* \to q + g/\gamma$ (1-loop) & QCD, QED, QCD$\otimes$QED \\
& $g/\gamma + \gamma^* \to q + \bar{q}$ (1-loop) & QCD, QED, QCD$\otimes$QED \\
& $q + \gamma^* \to q + g(\gamma) + g(\gamma)$ & QCD (QED) \\
& $q + \gamma^* \to q + g + \gamma$ & QCD$\otimes$QED \\
& $q + \gamma^* \to   q + q + \bar{q}$ & QCD, QED, QCD$\otimes$QED \\
& $q + \gamma^* \to   q + q' + \bar{q}'$ & QCD, QED \\
& $g(\gamma) + \gamma^* \to q + \bar{q} + g(\gamma)$ & QCD (QED) \\
& $g(\gamma) + \gamma^* \to q + \bar{q} + \gamma(g)$ & QCD$\otimes$QED \\
\hline
\end{tabular}
\caption{Partonic subprocesses contributing at NLO, and NNLO, classified by contribution type. 
All the partons in final state of process can hadronize, here $q'$ is a quark of flavor different from $q$.}
\label{tab:subprocesses}
\end{table}

The evaluation of phase-space integrals here is more involved than for inclusive DIS cross sections due to the kinematic constraint 
$\frac{z}{z_1} = \frac{p_a \cdot p_b}{p_a \cdot q}$.  
This constraint is imposed via a delta function replaced by a cut propagator using the reverse unitarity method~\cite{Anastasiou:2003gr,Anastasiou:2012kq}, which enables the use of integration-by-parts (IBP) identities~\cite{Chetyrkin:1981qh} to reduce integrals to a minimal set of master integrals (MIs).
The IBP reduction is carried out using the \texttt{LiteRed} package~\cite{Lee:2013mka}.  
The master integrals (MIs) needed for the pure QED and mixed QCD$\otimes$QED contributions coincide with those already known from pure QCD computations.
Details of the computation have been described in previous work~\cite{Ahmed:2024owh,Goyal:2024emo,Bonino:2024adk}.

After including the MIs results in the partonic cross sections, soft divergences cancel between real and virtual contributions and collinear divergences are removed by mass factorization using space- and time-like AP kernels. We apply eq.~(\ref{eq:massfactse}) to systematically extract the collinear finite CFs, $(\Delta)\mathcal{C}_{I,{ab}}$, order by order from the partonic cross sections. 
The AP kernels can be expanded as a power series in the strong coupling $a_s$ and the electromagnetic coupling $a_e$ as follows:
\begin{align}
\label{eq:APkernelsse}
\BbbGamma_{c\la d} &=\delta_{cd}\delta(1-\xi) 
 + \! \! \!\sum_{i+j=1}^{\infty}a_s^i(\mu_F^2)a_e^j(\mu_F^2)~\BbbGamma^{(i,j)}_{c\la d}(\xi,\varepsilon)
\, ,
\end{align}
where we abbreviate the AP kernels collectively by the set $\BbbGamma \in \{ \Gamma,\Delta \Gamma, \tilde{\Gamma} \}$.
They satisfy renormalization group evolution equations,
$\mu_F^2 \frac{d }{d\mu_F^2}\BbbGamma_{c\la d} = \frac{1}{2}
\mathds{P}_{c k}(a_s(\mu_F^2),a_e(\mu_F^2))\otimes \BbbGamma_{k\la d}$
in terms of splitting functions, denoted by the set $\mathds{P} \in \{ \mathrm{P},\Delta\mathrm{P}, \mathrm{\tilde{P}}\}$ 
consisting of 
unpolarized space-like ($\mathrm{P}$), 
polarized space-like ($\Delta\mathrm{P}$) and 
unpolarized time-like ($\mathrm{\tilde{P}}$) splitting functions respectively.  
Their perturbative expansion reads $\mathds{P}_{ab} =  \sum_{i+j=1}^{\infty} a_s^{i}a_e^{j} \mathds{P}_{ab}^{(i,j)}$ where $i$ and  $j$ are the QCD and QED indices,  respectively. 
For removal of initial state collinear singularities in the present study we use the QCD space-like (polarized) unpolarized splitting functions $(\Delta)\mathrm{P}_{ab}^{(i,0)}$ with $i=1,2$ ~\cite{Moch:2004pa,Vogt:2004mw,
Mertig:1995ny,Vogelsang:1995vh,Vogelsang:1996im,Moch:2014sna,Blumlein:2021enk,Blumlein:2021ryt,Blumlein:2022gpp,Almasy:2011eq,Chen:2020uvt}, 
 the space-like splitting functions $(\Delta)\mathrm{P}_{ab}^{(0,j)}$ with $j=1,2$ for QED and $(\Delta)\mathrm{P}_{ab}^{(1,1)}$ for the mixed QCD$\otimes$QED, all available in the literature~\cite{deFlorian:2015ujt,deFlorian:2016gvk,deFlorian:2025yar}. 
The time-like splitting functions needed to remove final state collinear singularities are known for QCD interactions~\cite{Almasy:2011eq,Chen:2020uvt}.
For QED and the mixed QCD$\otimes$QED interactions, the corresponding expressions are not available in the literature. 
We derive them here by assuming that mass factorization holds with QED interactions and by requiring finite CFs. 
In detail, we determine the time-like splitting functions $\tilde{\mathrm{P}}_{ab}^{(0,j)}, j=1,2$ for QED  and 
$\tilde{\mathrm{P}}_{ab}^{(1,1)}$ for mixed QCD$\otimes$QED :  
\begin{align}
\rm {LO} &: \tilde{\mathrm{P}}_{qq}^{(0,1)},\tilde{\mathrm{P}}_{q\gamma}^{(0,1)},\tilde{\mathrm{P}}_{\gamma q}^{(0,1)},\tilde{\mathrm{P}}_{\gamma \gamma}^{(0,1)} \nonumber \\
\rm {NLO} &:\tilde{\mathrm{P}}_{\gamma q}^{(0,2)},\tilde{\mathrm{P}}_{g q}^{(1,1)},\tilde{\mathrm{P}}_{\gamma q}^{(1,1)},\tilde{\mathrm{P}}_{q q}^{(0,2),\mathrm{NS}}, 
\tilde{\mathrm{P}}_{q q}^{(0,2),\mathrm{S}},\tilde{\mathrm{P}}_{q \overline{q}}^{(0,2)}  
\end{align}
The QCD CFs for the polarized SF $g_1$ are already known \cite{Bonino:2024wgg,Goyal:2024tmo,Bonino:2025bqa}. 
As mentioned before, we use Larin's prescription for $\gamma_5$, with polarized space-like splitting functions in Larin's scheme and unpolarized time-like splitting functions in $\overline{\text{MS}}$ to obtain the QED and mixed QCD$\otimes$QED CFs. 
The CFs in the $\overline {\text{MS}}$-scheme are then derived with the help of a finite scheme transformation $Z = 1+ \sum_{i+j=1} a_s^i a_e^j Z^{(i,j)}$.  
The ones for QCD, $Z^{(i,0)}$ are already known for $i=1,2$. 
The remaining  $Z^{(i,j)}$ i.e. $(i,j)=(0,1),(0,2),(1,1)$ relevant  for QED and the mixed QCD$\otimes$QED CFs are obtained from the former through Abelianisation, see, e.g.,~\cite{deFlorian:2015ujt}. 
The set of polarized $\overline{\text{MS}}$ splitting functions given below for QED and mixed QCD$\otimes$QED evolution, that has been obtained in this way, i.e., using $Z$ for $g_1$, is in complete agreement with \cite{deFlorian:2023zkc,deFlorian:2025yar}.
\begin{eqnarray*}
{\rm LO} &:& \Delta{\mathrm{P}}_{qq}^{(0,1)},\Delta{\mathrm{P}}_{q\gamma}^{(0,1)},\Delta{\mathrm{P}}_{\gamma q}^{(0,1)},\Delta{\mathrm{P}}_{\gamma \gamma}^{(0,1)}  \nonumber \\
{\rm NLO} &:&
\Delta{\mathrm{P}}_{q\gamma }^{(0,2)},\Delta{\mathrm{P}}_{qg}^{(1,1)},
  \Delta{\mathrm{P}}_{q \gamma}^{(1,1)}, 
\Delta{\mathrm{P}}_{q q}^{(0,2),\mathrm{NS}}, \nonumber \\ && 
 \Delta{\mathrm{P}}_{q q}^{(0,2),\mathrm{S}},\Delta{\mathrm{P}}_{q \overline{q}}^{(0,2)} 
\end{eqnarray*}
where the superscripts $\mathrm{NS}$ ($\mathrm{S}$) refer to the non-singlet (singlet) parts of the $qq$ channel, respectively. 
This completes the NNLO derivation of the CFs via eq.~(\ref{eq:massfactse}); as in the known QCD case, the finite CFs can be expressed in terms of double and single distributions, as well as regular terms, cf.~\cite{Goyal:2024emo} for computational details.   
The convolution of the CFs with PDFs and FFs provides $(g_1)F_1 = \sum_{i+j=0} a_s^ia_e^j (g_1)F_1^{(i,j)}$, such that at LO $F_1^{(0,0)}=\sum_{q} e_{q}^{2} (\Delta)H_{qq}$ ($e_q$ being the electric charge of quark $q$). Since we have already presented the QCD results for $F_i$ and $g_1$, that is,  $(g_1) F_I^{(i,0)}$ for $i=1,2$  in  \cite{Goyal:2024emo},
in the following we give results for pure QED and mixed QCD$\otimes$QED:
\begin{widetext}
\begin{align}
\label{eq:nlo-F1g1}
(g_1)F_1^{(0,1)} & = \sum_{q} e_{q}^{4}\bigg(   (\Delta)H_{qq}\hat \otimes (\Delta)\mathcal{C}_{1,qq}^{(0,1)} 
+  (\Delta)H_{q\gamma} \hat \otimes (\Delta)\mathcal{C}_{1,q\gamma}^{(0,1)} +  (\Delta)H_{\gamma q} \hat \otimes (\Delta)\mathcal{C}_{1,\gamma q}^{(0,1)} \bigg) 
\, ,
\end{align}

\begin{align}
\label{eq:nnlo-F1g1}
(g_1)F_1^{(0,2)} &= \sum_q e_q^{6}\bigg(  
   (\Delta)H_{qq} \hat\otimes (\Delta)\mathcal{C}_{1,qq,[1]}^{(0,2)}  +  (\Delta)H_{q\bar{q}} \hat\otimes (\Delta)\mathcal{C}_{1,q\bar{q}}^{(0,2)}
+  (\Delta)H_{q\gamma} \hat\otimes (\Delta)\mathcal{C}_{1,q\gamma}^{(0,2)}
+  (\Delta)H_{\gamma q} \hat\otimes (\Delta)\mathcal{C}_{1,\gamma q}^{(0,2)}
\bigg) \nonumber \\ 
&
+ \sum_q e_q^{4}\Big(N\sum_{q_i} e_{q_i}^{2} + \sum_{l_i} e_{l_i}^{2}  \Big)   (\Delta)H_{qq} \hat\otimes (\Delta)\mathcal{C}_{1,qq,[2]}^{(0,2)} + \sum_q e_q^{2}\Big(N\sum_{q_i} e_{q_i}^{4} + \sum_{l_i} e_{l_i}^{4}  \Big)
   (\Delta)H_{qq} \hat\otimes (\Delta)\mathcal{C}_{1,qq,[3]}^{(0,2)} \nonumber\\
   &
+ \sum_q \sum_{q' \ne q} \bigg(
  e_q^{4}e_{q'}^{2}   (\Delta)H_{qq'}^+ \hat\otimes (\Delta)\mathcal{C}_{1,qq',[1]}^{(0,2)}
+ e_{q}^{2}e_{q'}^{4}  (\Delta)H_{qq'}^+ \hat\otimes (\Delta)\mathcal{C}_{1,qq',[2]}^{(0,2)}
+ e_q^{3} e_{q'}^{3}  (\Delta)H_{qq'}^- \hat\otimes (\Delta)\mathcal{C}_{1,qq',[3]}^{(0,2)}
\bigg)
\nonumber\\
   &
+ \Big( N\sum_{q_i} e_{q_i}^{6} +\sum_{l_i} e_{l_i}^{6} \Big) \bigg(
 (\Delta)H_{\gamma\gamma} \hat\otimes (\Delta)\mathcal{C}_{1,\gamma\gamma}^{(0,2)}
\bigg) 
,
\end{align}
\begin{align}
\label{eq:nnlo-F1g1(1,1)}
(g_1)F_1^{(1,1)} &= \sum_{q} e_{q}^{4}\bigg( (\Delta)H_{qq}\hat \otimes (\Delta)\mathcal{C}_{1,qq}^{(1,1)} +  (\Delta)H_{q\bar{q}}\hat \otimes  (\Delta)\mathcal{C}_{1,q\bar{q}}^{(1,1)} +  (\Delta)H_{qg} \hat \otimes (\Delta)\mathcal{C}_{1,qg}^{(1,1)} +  (\Delta)H_{g q} \hat \otimes (\Delta)\mathcal{C}_{1,g q}^{(1,1)}
\nonumber\\
&
+  (\Delta)H_{q\gamma} \hat \otimes (\Delta)\mathcal{C}_{1,q\gamma}^{(1,1)} 
+  (\Delta)H_{\gamma q} \hat \otimes (\Delta)\mathcal{C}_{1,\gamma q}^{(1,1)} \bigg) + \Big( \sum_{q_i} e_{q_i}^{4} \Big) \bigg(
(\Delta)H_{g\gamma} \hat \otimes (\Delta)\mathcal{C}_{1,g\gamma}^{(1,1)} 
+  (\Delta)H_{\gamma g} \hat \otimes (\Delta)\mathcal{C}_{1,\gamma g}^{(1,1)} \bigg)   
\, ,
\end{align}
where $e_{q'}$ 
is the electric charge of quark $(q')$ of different
flavor from quark $q$ and $\hat\otimes$ denotes the convolution with $ (\Delta)H_{ab}$ in both variables $x$ and $z$,
\begin{align}
 (\Delta)H_{qq} &=  (\Delta)f_q(x) D_q(z) +  (\Delta)f_{\bar{q}} (x)D_{\bar{q}}(z)\, , 
\hspace{1.5cm} (\Delta)H_{q\bar{q}} =  (\Delta)f_q (x) D_{\bar{q}} (z)+  (\Delta)f_{\bar{q}}(x) D_q(z)\, , 
\nonumber\\
 (\Delta)H_{qg} &=  (\Delta)f_q (x) D_g (z)+ (\Delta)f_{\bar{q}}(x) D_g (z)\, , 
\hspace{1.5cm} (\Delta)H_{q\gamma} =  (\Delta)f_q (x) D_\gamma (z)+  (\Delta)f_{\bar{q}}(x) D_\gamma (z)\, , 
\nonumber\\
(\Delta)H_{\gamma q} &=   (\Delta)f_\gamma (x) D_q (z) +    (\Delta)f_\gamma(x) D_{\bar{q}}(z)\, , 
\hspace{1.5cm}(\Delta)H_{gq} =   (\Delta)f_g (x) D_q (z) +    (\Delta)f_g(x) D_{\bar{q}}(z)\, , 
 \nonumber\\
 (\Delta)H_{gg} &=   (\Delta)f_g (x)D_g(z)\, , 
\hspace{0.2cm}(\Delta)H_{\gamma \gamma} =   (\Delta)f_\gamma (x)D_\gamma(z)\, ,
\hspace{0.2cm} (\Delta)H_{g \gamma} =   (\Delta)f_g (x) D_\gamma (z) \,   ,
\hspace{0.2cm} (\Delta)H_{\gamma g} =   (\Delta)f_\gamma (x) D_g (z) \, ,
\nonumber \\
 (\Delta)H^{\pm}_{qq'} &=   (\Delta)f_q (x)D_{q'}(z) \pm   (\Delta)f_q (x)D_{\bar{q}'}(z) 
\pm    (\Delta)f_{\bar{q}} (x)D_{q'}(z)+    (\Delta)f_{\bar{q}} (x) D_{\bar{q}'}(z)\, .
\end{align}
\end{widetext} 

\onecolumngrid

\begin{figure*}[ht]
\vspace*{2mm}
\includegraphics[width=0.495\textwidth]{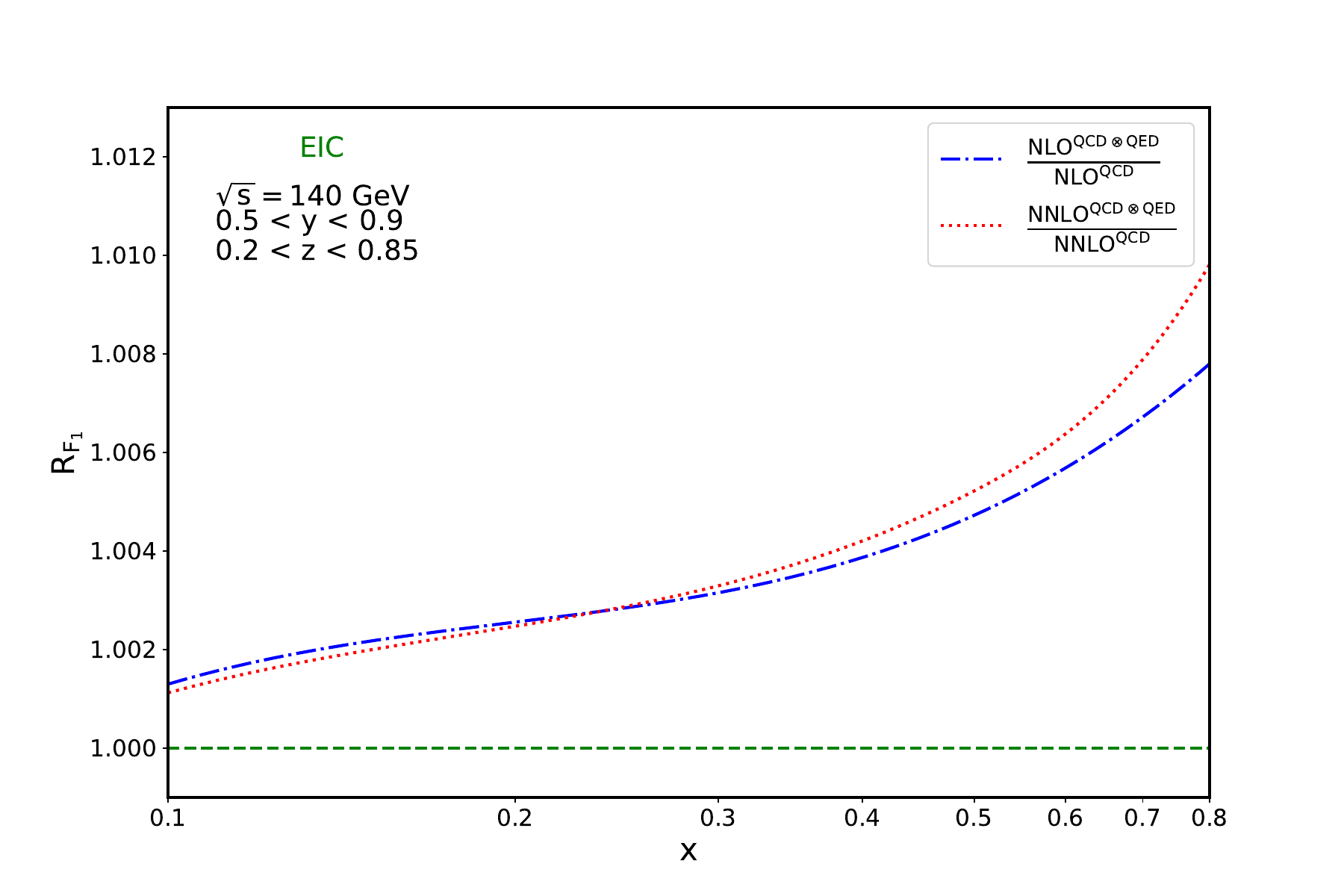}
\includegraphics[width=0.495\textwidth]{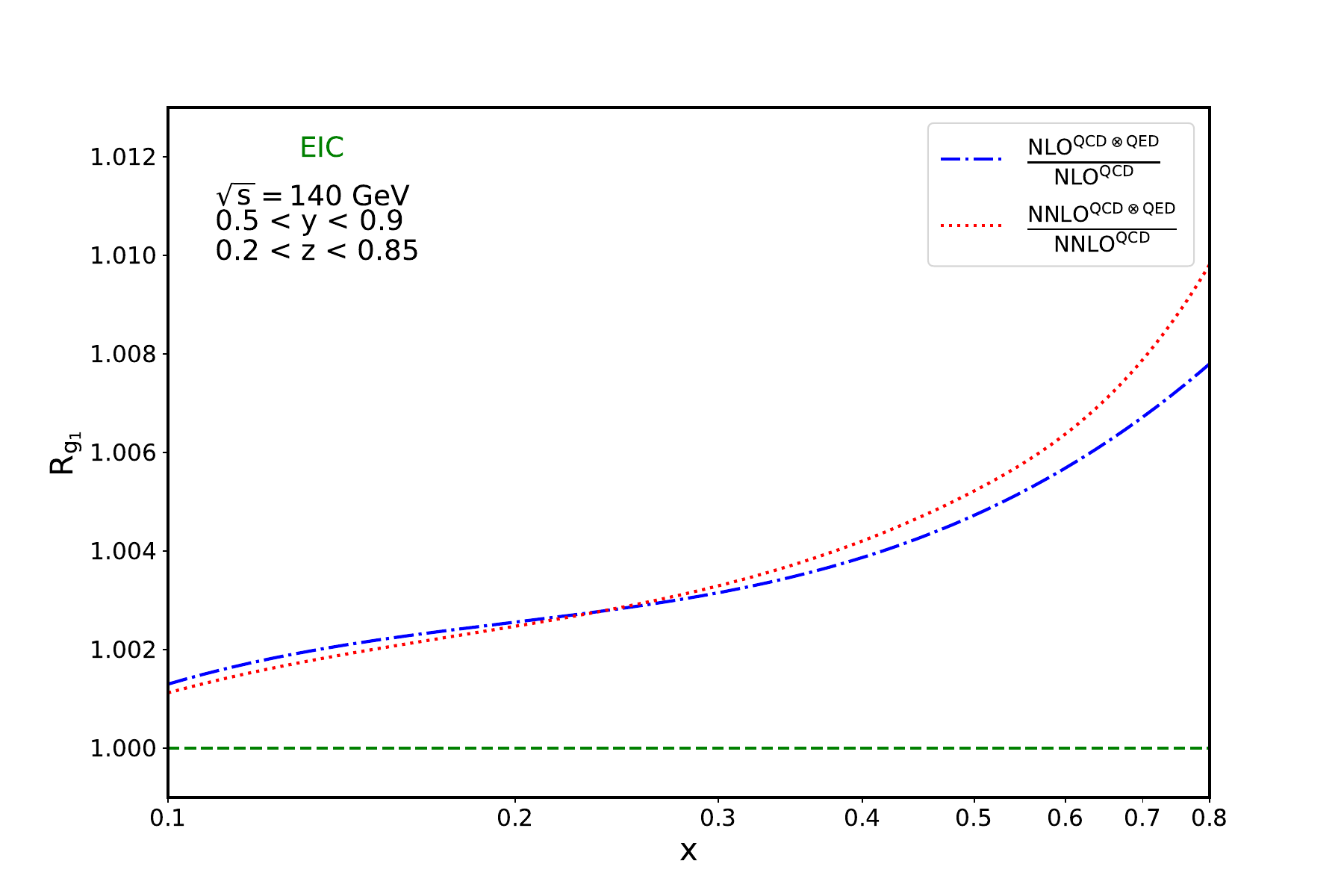}
\caption{Ratio of the SFs $\text{F}_1$ (left panel) and $g_1$ (right panel) with QCD$\otimes$QED contributions at NLO and NNLO to those with only QCD corrections applied at the respective order as a function of $x$ at the central scale $\mu_R^2$ = $\mu_F^2$ = $Q^2_{\text{avg}}$ for the EIC at $\sqrt{s}=140$ GeV.
Integration ranges for $y$ and $z$ are indicated in the plots.}
\label{fig:1FGz} 
\end{figure*}

\begin{figure*}[ht]
\includegraphics[width=0.495\textwidth]{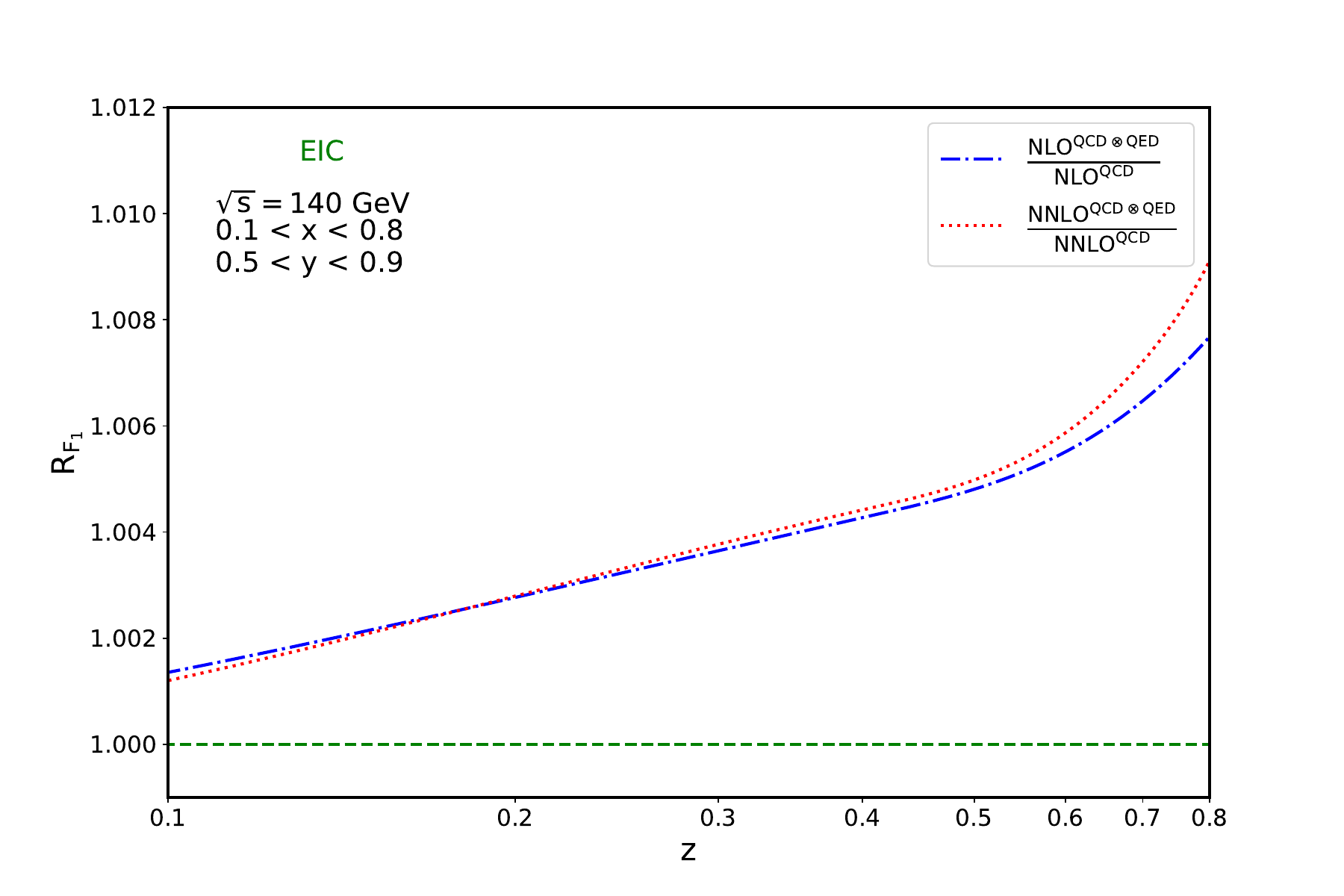}
\includegraphics[width=0.495\textwidth]{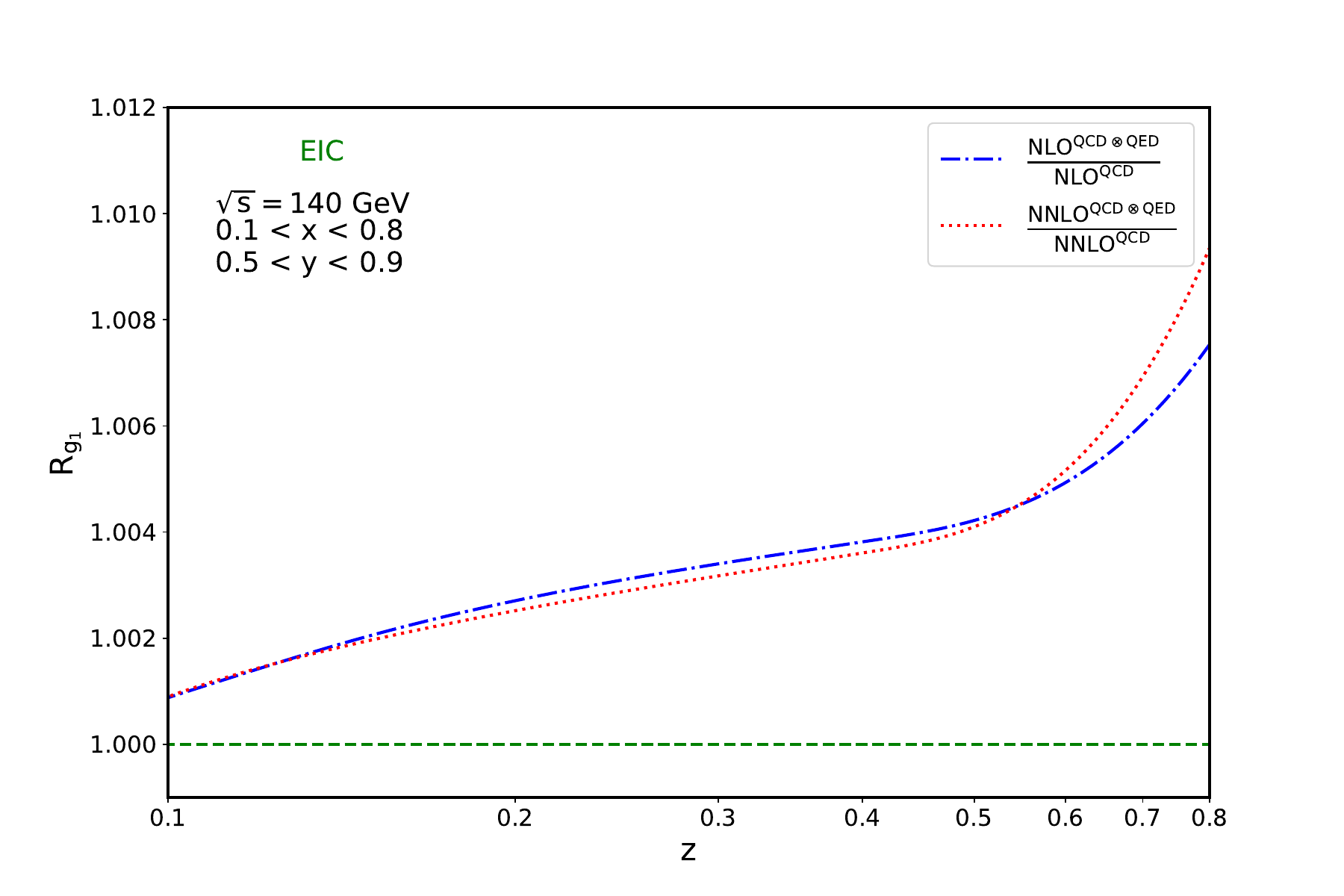}
\caption{Same as Fig.~\ref{fig:1FGz} for the SFs $F_1$ (left panel) and $g_1$ (right panel) as a functions of $z$ and integration ranges for $x$ and $y$ indicated in the plots.}
\label{fig:1FGx}
\end{figure*}

\renewcommand{\arraystretch}{1.2} 
\begin{table*}[!hbt]
\vspace*{2mm}
\begin{tabular}
{|p{1.0cm}|p{2.5cm}|p{2.5cm}|p{2.5cm}|p{2.5cm}|p{2.5cm}|}
\hline
\centering
  & \multicolumn{4}{c}{\hspace{2.0cm}$F_1$} & \\
\cline{2-4} \cline{5-6}
\centering
  $z$ &\vspace{0.015cm} LO & \vspace{0.015cm} $\rm NLO^{QCD}$  &\vspace{0.015cm} $\rm NLO^{QCD\otimes QED}$  &\vspace{0.015cm} $\rm NNLO^{QCD}$ & \vspace{0.015cm} $\rm NNLO^{QCD \otimes QED}$ \\
\hline
 $0.15$  & 
 $1.23119^{+4.644\%}_{-4.288\%}$ & 
 $1.04009^{+0.952\%}_{-1.202\%}$ &
 $1.04266^{+0.865\%}_{-1.121\%}$ &
 $0.99712^{+0.092\%}_{-0.176\%}$ & 
 $0.99894^{+0.175\%}_{-0.077\%}$   \\
 \hline
 $0.25$   & 
 $0.44911^{+5.814\%}_{-5.283\%}$ & 
 $0.38139^{+1.341\%}_{-1.620\%}$ &
 $0.38257^{+1.245\%}_{-1.532\%}$ & 
 $0.33861^{+0.156\%}_{-0.291\%}$ & 
 $0.33962^{+0.156\%}_{-0.211\%}$   \\
  \hline
$0.45$ & 
$0.08062^{+7.564\%}_{-6.741\%}$ & 
$0.07737^{+2.583\%}_{-2.825\%}$ & 
$0.07771^{+2.474\%}_{-2.727\%}$ & 
$0.07296^{+0.433\%}_{-0.720\%}$ & 
$0.07329^{+0.393\%}_{-0.597\%}$     \\
  \hline
 $0.6$   & 
 $0.02339^{+8.625\%}_{-7.611\%}$ & 
 $0.02469^{+3.588\%}_{-3.767\%}$ & 
 $0.02483^{+3.471\%}_{-3.666\%}$ & 
 $0.02463^{+1.031\%}_{-1.372\%}$ & 
 $0.02479^{+0.885\%}_{-1.234\%}$    \\
  \hline
 $0.75$   & 
 $0.00651^{+10.229\%}_{-8.902\%}$ & 
 $0.00682^{+5.019\%}_{-5.069\%}$  &
 $0.00687^{+4.894\%}_{-4.964\%}$  & 
 $0.00699^{+1.897\%}_{-2.294\%}$ &
 $0.00705^{+1.731\%}_{-2.141\%}$    \\
  \hline\hline  
\end{tabular}
\caption{
Values of the SF $F_1$ at various orders at the central scale $\mu_R^2$ = $\mu_F^2$ = $Q^2_{\text{avg}}$ and variation of scales \{$\mu_R^2$,$\mu_F^2$\} $\in$ $\left[Q^2_{\text{avg}}/2,2Q^2_{\text{avg}}\right]$ as function of $z$. (See text for integration ranges of $x$, $y$).}
\label{tab:F1_z}
\end{table*}

\begin{table*}[!hbt]
\vspace*{2mm}
\begin{tabular}{|p{1.0cm}|p{2.5cm}|p{2.5cm}|p{2.5cm}|p{2.5cm}|p{2.5cm}|}
\hline
\centering
  & \multicolumn{4}{c}{\hspace{2.0cm}$g_1$} & \\
\cline{2-4} \cline{5-6}
\centering
  $z$ &\vspace{0.015cm} LO & \vspace{0.015cm} $\rm NLO^{QCD}$  &\vspace{0.015cm} $\rm NLO^{QCD\otimes QED}$  &\vspace{0.015cm} $\rm NNLO^{QCD}$ & \vspace{0.015cm} $\rm NNLO^{QCD\otimes QED}$ \\
\hline
 $0.15$  & 
 $0.59417^{+4.356\%}_{-4.034\%}$ & 
 $0.45428^{+0.471\%}_{-0.741\%}$ &
 $0.45525^{+0.467\%}_{-0.653\%}$ &
 $0.42511^{+0.142\%}_{-0.183\%}$ & 
 $0.42599^{+0.237\%}_{-0.278\%}$  \\
 \hline
 $0.25$   & 
 $0.22688^{+5.579\%}_{-5.080\%}$ & 
 $0.17305^{+0.804\%}_{-1.121\%}$ &
 $0.17353^{+0.702\%}_{-1.026\%}$ &
 $0.14799^{+0.154\%}_{-0.158\%}$ & 
 $0.14837^{+0.250\%}_{-0.268\%}$  \\
  \hline
$0.45$ & 
 $0.04229^{+7.416\%}_{-6.614\%}$ & 
 $0.03790^{+2.001\%}_{-2.313\%}$ &
 $0.03806^{+1.888\%}_{-2.212\%}$ &
 $0.03430^{+0.266\%}_{-0.373\%}$ & 
 $0.03443^{+0.305\%}_{-0.286\%}$  \\
  \hline
 $0.6$   & 
 $0.01232^{+8.500\%}_{-7.505\%}$ & 
 $0.01236^{+3.017\%}_{-3.284\%}$ &
 $0.01242^{+2.896\%}_{-3.178\%}$ &
 $0.01192^{+0.628\%}_{-0.990\%}$ & 
 $0.01199^{+0.558\%}_{-0.845\%}$  \\
  \hline
 $0.75$   & 
 $0.00341^{+10.116\%}_{-8.809\%}$ & 
 $0.00339^{+4.422\%}_{-4.583\%}$ &
 $0.00342^{+4.293\%}_{-4.473\%}$ &
 $0.00337^{+1.414\%}_{-1.856\%}$ & 
 $0.00340^{+1.245\%}_{-1.698\%}$  \\
  \hline\hline  
\end{tabular}
\caption{
Same as Tab.~\ref{tab:F1_z} for the SF $g_1$.}
\label{tab:g1_z}
\end{table*}

\twocolumngrid

We now illustrate the numerical impact of our  $(\Delta){\mathcal C}_{1,ab}$ results  for various centre-of-mass energies $\sqrt{s}$ for a range of $x$ and $z$ values. 
Specifically, we examine the effects of pure QED and mixed QCD$\otimes$QED corrections on the SFs $F_1(g_1)$, 
included alongside the pure QCD terms, providing a more complete description up to $\mathcal{O}(\alpha_e\,\alpha_s)$.  
In Fig.~\ref{fig:1FGz}, the left (right) panel shows results for $F_1$ ($g_1$) as a function of $x$, after integration of $y$ between $0.5$ and $0.9$ and of $z$ between $0.2$ and $0.85$. 
We have set both renormalisation and factorization scale equal to central scale, $\mu_F = \mu_R = Q_{\rm avg}$, where $Q^2_{\rm avg} = x y_{\rm avg} s$.
Similarly, in Fig.~\ref{fig:1FGx}, the left (right) panel presents $F_1$ ($g_1$) as a function of $z$, after integration of $x$ between $0.1$ and $0.8$ and of $y$ between $0.5$ and $0.9$,  we set $\mu_F = \mu_R = Q_{\rm avg}$ here, $Q^2_{\rm avg} = x_{\rm avg} y_{\rm avg} s$. We have used $n_f$ = 3 as active number of quark and lepton flavors for all numerical calculations.
In table \ref{tab:F1_z} (\ref{tab:g1_z}), we present the results of $F_1$ ($g_1$) from pure QCD and mixed QCD$\otimes$QED for $z=0.15,0.25,0.45,0.6,0.75$ along with percentage error coming from varying $\mu_R$ and $\mu_F$.  
Both in the plots and in the tables, NLO$^{\text{QCD}\otimes \text{QED}}$ indicates that contributions at order $a_s$ from QCD and at $a_e$ from QED are  added to LO; in NNLO$^{\text{QCD}\otimes \text{QED}}$ we have added to LO the pure QCD
contributions upto order $a_s^2$, pure QED to order $a_e^2$ and mixed QCD$\otimes$QED upto $a_s a_e$.
For $F_1$, we have used the \texttt{LUXqed} PDF sets~\cite{NNPDF:2017mvq} at LO, NLO and NNLO, extended to include QED effects. 
In contrast, the predictions for $g_1$ have been obtained using 
the \texttt{BDSSV24NLO} PDF set at LO and NLO, and the \texttt{BDSSV24NNLO} PDF set~\cite{Borsa:2024mss} at the NNLO level, also incorporating QED effects. 
In both cases, for $F_1$ and $g_1$, the \texttt{NNFF10PIp} FF sets~\cite{Bertone:2017tyb} have been utilized at the respective orders (LO, NLO, NNLO). 
The contribution coming from QED and mixed QCD$\otimes$QED is comparable in size to the \text{N$^{3}$LO} QCD corrections.  
Since the polarized PDFs and unpolarized FFs for photons are not available in the literature, we have put them to be zero throughout. 
We observe that the dominant contribution come from the QCD sector while those from QED and mixed QCD$\otimes$QED are generally small but not negligible.  However, the later is expected to be comparable in size to the \text{N$^{3}$LO} QCD corrections.

In this letter, we present the complete NNLO coefficient functions for both unpolarized and polarized SIDIS in the \(\overline{\text{MS}}\)-scheme, including pure QED and mixed QCD$\otimes$QED corrections. 
We also provide the relevant time-like unpolarized splitting functions, the space-like polarized splitting functions in \(\overline{\text{MS}}\)-schemes, and the renormalization constant $Z$ required for the scheme transformation at NNLO. 
These results address a key gap in the literature and facilitate high-precision predictions for SIDIS, supporting future global fits of (un)polarized PDFs and unpolarized FFs.
In conclusion, our results underscore the importance of including QED and mixed 
QCD$\otimes$QED corrections in precision studies, as they enhance the robustness and accuracy of theoretical predictions at the EIC.

Supplementary files are provided with this work, including \texttt{Mathematica} notebooks containing the
unpolarized CFs, ${\mathcal{C}}_{I,ab}^{(i,j)}$
and polarized CF, ${\Delta\mathcal{C}}_{1,ab}^{(i,j)}$ in the $\overline{\text{MS}}$ scheme, the required splitting functions and renormalization constant for scheme transformation $Z$. All CFs are simplified to eliminate the $\theta$ functions making all the  terms real-valued and continuous throughout the entire range 
$0 \le x', z' \le 1$.

%
%
\section*{Acknowledgements} 
We would like to thank N. Rana for fruitful discussions.
This work has been supported through a joint Indo-German research grant by
the Department of Science and Technology (DST/INT/DFG/P-03/2021/dtd.12.11.21). 
S.M. acknowledges the ERC Advanced Grant 101095857 {\it Conformal-EIC}. 
In addition we would also like to thank the computer administrative unit of IMSc for their help and support.

\bibliographystyle{apsrev4-1}
\bibliography{main}

\end{document}